\documentclass[aps,showpacs,twocolumn]{revtex4}

\usepackage{graphicx}
\usepackage[dvipdfm]{color}
\usepackage{amsmath}
\usepackage{amssymb}
\usepackage{mathrsfs}

\begin{document}



\title{
$N+1$ formalism in Einstein-Gauss-Bonnet gravity
} 

\author{
Takashi Torii$^{(1)}$
}
\email{torii@ge.oit.ac.jp}
\author{
Hisa-aki Shinkai$^{(2)}$
}
\email{shinkai@is.oit.ac.jp}

\address{ 
$^{(1)}\!\!$
Department of General Education, 
Osaka Institute of Technology,
Omiya, Asahi-ku, Osaka 535-8585, Japan
}

\address{
$^{(2)}\!\!$
Department of Information Systems, 
Osaka Institute of Technology, 
Kitayama, Hirakata, Osaka 573-0196, Japan
}

\date{\today}

\begin{abstract}
Towards the investigation of the full dynamics in higher-dimensional and/or
stringy gravitational model, we present 
the basic equations of the Einstein-Gauss-Bonnet gravity theory.
We show $(N+1)$-dimensional version of the ADM decomposition
including Gauss-Bonnet terms, which shall be the standard approach 
to treat the space-time as a Cauchy problem.  
Due to the quasi-linear property of the Gauss-Bonnet gravity, we find that 
the evolution equations can be in a treatable form in numerics.  
We also show the conformally-transformed constraint equations 
for constructing an initial data.  We discuss how the constraints can be 
simplified by tuning the powers of conformal factors.  
Our equations can be used both for timelike and spacelike foliations. 
\end{abstract}




\maketitle

\section{Introduction}
\label{sec:Introduction}

General relativity (GR) has been tested with many experiments and observations
both in the strong and weak gravitational field regimes (see e.g. \cite{will}), 
and none of them are contradictory to GR.  However, the theory also predicts the
appearance of the spacetime singularities under natural conditions \cite{Hawking,CCH}, which also indicates that GR is still incomplete as a physics theory
that describes whole of the gravity and the spacetime 
structure.

We expect that the true fundamental theory will resolve these theoretical problems. Up to now, several quantum theories of gravity have been proposed. Among them superstring/M-theory, formulated in higher dimensional spacetime, is the most promising candidate. In these prospect, a considerable number of studies concerned with gravitational phenomena and cosmology have been made in the string theoretical framework beyond GR.

Since the present knowledge is still far from understanding the full aspects of the string theory, several kinds of approaches, which are fundamentally approximation, are usually taken. 
Among them perturbative approach plays important roles. There are two particular parameters which characterize the system in the superstring theory. One is string coupling parameter $g_s^2 = e^{\phi}$, where $\phi$ is the dilaton field. 
The other is the inverse string tension $\alpha'$. When the tension is strong (i.e., small $\alpha'$) compared to the energy scale of the system, it is difficult to excite strings and the size of the strings becomes small enough to be regarded as particles in the zeroth order approximation. In this limit 
GR (with other light fields) is recovered. This is called $\alpha'$-expansion\cite{Gross}.

In the higher order terms of $\alpha'$, curvature corrections appear. 
The Gauss-Bonnet (GB) term is the next leading order of the $\alpha'$-expansion of type IIB superstring theory\cite{Gross, Bento}, and has nice properties such that it is ghost-free combinations\cite{Zwiebach} and does not give higher derivative equations but an ordinary set of equations with up to second derivative in spite of the higher curvature combinations. 

The models with the GB term and/or other higher curvature terms have been intensively studied in the high energy physics. One of them is a series of studies in string cosmology. The pre-big bang scenario\cite{veneziano} is a fantastic scenario which tries to avoid the Big-Bang singularity by making use of T-duality\cite{S-dual} (or scale factor duality). 
Furthermore, the pre-big bang scenario gives a natural inflation mechanism, since the solution in the pre-big bang phase is inflating from the beginning at least in the string frame. 
Although these analyses show that the singularity problem has not been resolved yet completely, there are some cosmological solutions which do not start from an initial singularity\cite{Ishihara,Easther,Ohta}. 

In addition to cosmology, the string effects can be also seen in the study of black hole physics. As the size of a black hole becomes small, and the curvature around the black hole becomes large; it is expected that the curvature corrections cannot be negligible. 
The singularity inside of the event horizon would be also modified or even disappeared by string effects. For these reasons, the static or stationary black hole solutions in effective string theories were investigated both in the systems without higher curvature terms\cite{GM,GHS} and with such terms\cite{Boulware,Torii,Torii_2}. Besides the static or stationary solutions, there are some dynamical solutions which are motivated by gravitational collapse\cite{Maeda-H}.  

These analysis are performed on the assumption of highly symmetric 
spacetime because the system is much more complicated than that in 
GR. To obtain deeper understanding of the early stage of the universe, 
singularity, and/or black holes, we should consider less symmetric 
and/or dynamical spacetime; the analyses requires the direct numerical 
integration of the equations. None of the fully dynamical simulations 
in GB gravity has been performed.

In this article, we present the basic equations of the Einstein-GB gravity theory.
We show $(N+1)$-dimensional version of the ADM decomposition, which is the standard 
approach to treat the spacetime as a Cauchy problem. 
The topic was first discussed by Choquet-Bruhat \cite{choquetbruhat1988}, but 
the full set of equations and the methodology have not yet been presented. 
In 4-dimensional GR, 
numerical simulations of binary compact objects are available in the last years, 
and many groups apply the modified ADM equations in order to obtain long-term stable 
and accurate simulations.   
However, such modifications are depend on the problem to consider, and the 
``robustest" formulation is not yet known (see e.g. \cite{shinkai2008}). 
Therefore, as the first step, we in this paper 
just present the fundamental space-time decomposition of the GB equations, 
focusing on GB term. 

The ADM decomposition is supposed to construct the spacetime with foliations 
of the constant-time hypersurfaces.  This method can be also applied to study  
the brane-world model\cite{braneworld},  which states the visible space-time is embedded in higher dimensional ``bulk" spacetime. 
As was first investigated  by Chamblin et al \cite{chargedBH}, it is possible
to study the ``bulk" structure 
by switching the normal vector of the hypersurface from timelike to spacelike.
We therefore present all set of equations for both cases for future convenience.

The outline of this paper is as follows. In Sec.~\ref{sec:Equations}, we show that the set of equations are divided into two constraints and evolution equations along to the standard procedure. In Sec.~\ref{sec:constraints}, we present the conformal approach to solve the constraints which shall be used for preparing an initial data. 
In Sec.~\ref{sec:evolutions}, we show the dynamical equations in GR and in GB theory separately.
Sec.~\ref{sec:discussion} is devoted to the discussions and summary.
We think these expressions are useful for future dynamical investigations.

\section{$(N+1)$-decomposition in Einstein-Gauss-Bonnet gravity}
\label{sec:Equations}

\subsection{Model and Basic Equations} 

We start from the Einstein-Gauss-Bonnet action in 
$(N+1)$-dimensional spacetime $({\cal M}, g_{\mu\nu})$ 
which is described as
\footnote{The Greek indices move $0, 1, \cdots, N$, while the Latin indices move $1, \cdots, N$.
We follow the notations of \cite{MTW}.}:
\begin{eqnarray}
\label{bulk_action}
S =&& \!\!\!\!\!\!
\int_{\cal M} d^{N+1}X \sqrt{- g} \biggl[
{1 \over 2 \kappa^2} \left({\cal R} -  
 2 \Lambda +\alpha_{GB} {\cal L}_{GB}\right)
\nonumber \\
&& \;\;\;\;\;\;\;\;\;\;\;\;\;\;\;\;\;\;\;\;\;\;\;\;\;\;\;\;\;\;\;\;\;\;\;\;\;\;\;\;
+{\cal L}_{\rm matter} \biggr],
\end{eqnarray}
with
\begin{eqnarray}
{\cal L}_{GB}={\cal R}^2-4{\cal R}_{\mu\nu}{\cal R}^{\mu\nu}
+{\cal R}_{\mu\nu\rho\sigma}{\cal R}^{\mu\nu\rho\sigma}, 
\end{eqnarray}
where $\kappa^2$ is the $(N+1)$-dimensional gravitational constant, ${\cal R}$,
${\cal R}_{\mu\nu}$, ${\cal R}_{\mu\nu\rho\sigma}$ and ${\cal L}_{\rm matter}$ are the $(N+1)$-dimensional scalar curvature, Ricci tensor, Riemann  curvature and the matter Lagrangian, respectively. This action reproduces the standard $(N+1)$-dimensional Einstein gravity, if we set the coupling constant $\alpha_{GB} \: (\geq 0)$ equals to zero. 

The action (\ref{bulk_action}) gives the gravitational equation as
\begin{eqnarray}
{\cal G}_{\mu\nu}+ \alpha_{GB} {\cal H}_{\mu\nu}= \kappa^2 \,{\cal T}_{\mu\nu} \, ,
\label{Einstein}
\end{eqnarray}
where
\begin{eqnarray}
{\cal G}_{\mu\nu}=&& \!\!\!\!\!{\cal R}_{\mu\nu}-{1\over 2}g_{\mu\nu}{\cal R}
+g_{\mu\nu}\Lambda,
\\
{\cal H}_{\mu\nu}=&& \!\!\!\!\!
2\Bigl[{\cal R}{\cal R}_{\mu\nu}
-2{\cal R}_{\mu\alpha}{\cal R}^{\alpha}_{~\nu}
-2{\cal R}^{\alpha\beta}{\cal R}_{\mu\alpha\nu\beta}
\nonumber \\ 
&& \;\;\;\;\;\;\;
+{\cal R}_{\mu}^{~\alpha\beta\gamma}{\cal R}_{\nu\alpha\beta\gamma}\Bigr]
-{1\over 2}g_{\mu\nu}{\cal L}_{GB},
\label{EGB:eq}
\\ 
{\cal T}_{\mu\nu} =&& \!\!\!\!\!
-2 {\delta {\cal L}_{\rm matter} \over \delta
g^{\mu\nu}}  +g_{\mu\nu}{\cal L}_{\rm matter}.
\label{em_tensor_of_bulk}
\end{eqnarray}

\subsection{Projections to Hypersurface}
In order to investigate the space-time structure as the foliations of the
$N$-dimensional (spacelike or timelike) hypersurface $\Sigma$, 
we introduce the projection operator to $\Sigma$ as  
\begin{equation}
\bot_{\mu\nu}  
=g_{\mu\nu} -\varepsilon n_{\mu}n_{\nu},
\end{equation}
where $n_{\mu}$ is the unit-normal vector to $\Sigma$ with $n_{\mu}n^{\mu}=\varepsilon$, with which we define $n_\mu$ is timelike (if $\varepsilon=-1$)  or spacelike (if $\varepsilon=1$). Therefore, $\Sigma$ is spacelike (timelike) if $n_{\mu}$ is timelike (spacelike).

The projections of the gravitational equation (\ref{Einstein}) 
give the following three equations: 
\begin{eqnarray}
&& \!\!\!\!\!
\bigl({\cal G}_{\mu\nu}+\alpha_{GB}{\cal H}_{\mu\nu}\bigr) \, n^\mu \,  n^\nu=
\kappa^{2}  \,  {\cal T}_{\mu\nu} \,  n^ \mu \, n^\nu  = \kappa^{2} \rho,
\label{project1}
\\
&& \!\!\!\!\!
\bigl({\cal G}_{\mu\nu}+\alpha_{GB}{\cal H}_{\mu\nu}\bigr) \, n^\mu \, \bot_{\;\rho}^\nu
= 
\kappa^{2}   \,   {\cal T}_{\mu\nu} \, n^\mu\, \bot_\rho^\nu 
= - \kappa^{2}  J_\rho,
\label{project2} \rule[0mm]{0mm}{5mm}
\nonumber \\ 
\\
&& \!\!\!\!\!
\bigl({\cal G}_{\mu\nu}+\alpha_{GB}{\cal H}_{\mu\nu}\bigr) 
\, \bot_{\;\rho}^\mu  \, \bot_{\;\sigma}^\nu 
=\kappa^{2}  \,    {\cal T}_{\mu\nu}\, \bot_{\;\rho}^\mu  \, \bot_{\;\sigma}^\nu 
= \kappa^{2}  S_{\rho\sigma},
\label{project3} \rule[0mm]{0mm}{5mm}
\nonumber \\
\end{eqnarray}
where we defined the components of the energy-momentum tensor as 
\begin{equation}
 {\cal T}_{\mu\nu}=
\rho n_\mu n_\nu + J_\mu n_\nu +  J_\nu n_\mu + S_{\mu\nu},
\end{equation} 
and we also define $ {\cal T}= \varepsilon\rho + S^{\alpha}_{~\alpha}$ for later convenience. 

Projection of the $(N+1)$-dimensional Riemann tensor  onto the $N$-dimensional hypersurface can be written as 
\begin{eqnarray}
&&
{\cal R}_{\alpha\beta\gamma\delta} 
~\bot_{~\mu}^{\alpha} \, \bot_{~\nu}^{\beta} \, \bot_{~\rho}^{\gamma}\, \bot_{~\sigma}^{\delta}
\nonumber \\
&&
~~~~~~~~~=
R_{\mu\nu\rho\sigma}
-\varepsilon (K_{\mu\rho}K_{\nu\sigma}
-K_{\mu\sigma}K_{\nu\rho}),
\label{Riemann_}
\\
&&
{\cal R}_{\alpha\beta\gamma\delta} ~\bot_{~\mu}^{\alpha} \, \bot_{~\nu}^{\beta} \, \bot_{~\rho}^{\gamma}\, n^{\delta}
=-2D_{[\mu}K_{\nu]\rho},
\rule[0mm]{0mm}{6mm}
\label{Riemann_n} 
\\
&&
{\cal R}_{\alpha\beta\gamma\delta} 
~\bot_{~\mu}^{\alpha}\,  \bot_{~\rho}^{\gamma}\, n^{\beta}\, n^{\delta}
=\pounds_n K_{\mu\rho} +K_{\mu\alpha}\, K^{\alpha}_{~\rho}\, ,
\rule[0mm]{0mm}{6mm}
\label{Riemann_nn}
\end{eqnarray}
where  $R_{\mu\nu\rho\sigma}$ is the Riemann tensor of the induced metric 
$\gamma_{\mu\nu}(=\bot_{\mu\nu})$,
$D_{\mu}$ is the covariant differentiation with respect to $\gamma_{\mu\nu}$, 
$\pounds_n$ denotes the Lie derivative in the $n$-direction, and $K_{\mu\nu}$ is the extrinsic curvature defined as
\begin{eqnarray}
K_{\mu\nu}&=&-\frac12 \pounds_{n}\gamma_{\mu\nu}
=-\bot_{~\mu}^{\alpha} \bot_{~\nu}^{\beta} \nabla_{\alpha} n_{\beta}
\label{ex-curv}
\end{eqnarray}  
Eq.~(\ref{Riemann_}) is called the Gauss equation, and the contraction of (\ref{Riemann_n}) of $\mu$ and $\nu$ gives the Codacci equation. 

Using these projections, the $(N+1)$-dimensional Riemann curvature and
its contractions (the Ricci tensor and scalar curvature) are described by
the $N$-dimensional variables on the hypersurface $\Sigma$
as　
\begin{widetext}
\begin{eqnarray}
{\cal R}_{\mu\nu\rho\sigma}
&=&R_{\mu\nu\rho\sigma}
-\varepsilon \bigl(K_{\mu\rho}K_{\nu\sigma}-K_{\mu\sigma}K_{\nu\rho}
-n_{\mu} D_{\rho} K_{\sigma\nu}
+n_{\mu}D_{\sigma} K_{\rho\nu}
+n_{\nu}D_{\rho}K_{\sigma\mu}
-n_{\nu}D_{\sigma}K_{\rho\mu}
\nonumber
\\
&& ~~~
-n_{\rho} D_{\mu}K_{\nu\sigma}
+n_{\rho}D_{\nu}K_{\mu\sigma}
+n_{\sigma}D_{\mu}K_{\nu\rho}
-n_{\sigma}D_{\nu}K_{\mu\rho}\bigr)
+n_{\mu}n_{\rho} K_{\nu\alpha}K^{\alpha}_{~\sigma}
-n_{\mu}n_{\sigma} K_{\nu\alpha}K^{\alpha}_{~\rho}
\nonumber
\\
&& ~~~
-n_{\nu}n_{\rho} K_{\mu\alpha}K^{\alpha}_{~\sigma}
+n_{\nu}n_{\sigma} K_{\mu\alpha}K^{\alpha}_{~\rho}
+n_{\mu}n_{\rho} {\pounds_n} K_{\nu\sigma}
-n_{\mu}n_{\sigma} {\pounds_n} K_{\nu\rho}
\nonumber
\\
&& ~~~
-n_{\nu}n_{\rho} {\pounds_n} K_{\mu\sigma}
+n_{\nu}n_{\sigma} {\pounds_n} K_{\mu\rho},
\label{5DRiemann}
\\
\rule[0mm]{0mm}{8mm}
{\cal R}_{\mu\nu}
&=&R_{\mu\nu}-\varepsilon \Bigl[ KK_{\mu\nu}-2K_{\mu\alpha}K^{\alpha}_{~\nu}
+n_{\mu} \left(D_{\alpha} K^{\alpha}_{~\nu}- D_{\nu} K\right)
+n_{\nu} \left(D_{\alpha} K^{\alpha}_{~\mu}- D_{\mu} K\right) \Bigr]
\nonumber
\\
&& ~~~
+n_{\mu}n_{\nu} K_{\alpha\beta}K^{\alpha\beta}
+ \varepsilon {\pounds_n} K_{\mu\nu}
+n_{\mu}n_{\nu} \gamma^{\alpha\beta}{\pounds_n} K_{\alpha\beta},
\label{5DRicci}
\\
\rule[0mm]{0mm}{8mm}
{\cal R}
&=&R-\varepsilon \bigl(K^2-3K_{\alpha\beta}K^{\alpha\beta}
-2 \gamma^{\alpha\beta}{\pounds_n} K_{\alpha\beta}\bigr),
\label{5Dscalar}
\end{eqnarray}
where $K=K^{\alpha}_{~\alpha}$.

Substituting these relations into the field equation (\ref{Einstein}) or ~(\ref{project1})-(\ref{project3}), we find the equations are 
decomposed as 
\noindent
(a) the Hamiltonian constraint equation: 
\begin{equation}
M+\alpha_{GB}\bigl(M^2-4M_{ab}M^{ab}
+M_{abcd}M^{abcd}\bigr)
=-2\varepsilon\kappa^2\rho_H 
+2\Lambda 
\,,
\label{eq_Hamiltonian_const}
\end{equation}
(b) the momentum constraint equation: 
\begin{eqnarray}
&&
N_{i}+2\alpha_{GB} \bigl(MN_i
-2M_i^{\;a}N_a
+2M^{ab}N_{iab}
-M_{i}^{~cab}N_{abc}\bigr)
=\kappa^2 J_i
\,,\label{eq_momentum_const}
\end{eqnarray}
and (c) the evolution equations for $\gamma_{ij}$:
\begin{eqnarray}
&&
M_{ij}-{1\over2}M\gamma_{ij}
-\varepsilon \bigl(-K_{ia}K^a_{~j}+\gamma_{ij}K_{ab}K^{ab}
-{\pounds_n} K_{ij}+\gamma_{ij}\gamma^{ab}{\pounds_n} K_{ab}
\bigr)
\nonumber \\
&& ~~
+2\alpha_{GB}\Bigl[H_{ij}
+\varepsilon \bigl(M{\pounds_n} K_{ij} 
-2M^{\;a}_{i} {\pounds_n} K_{a j}
-2M^{\;a}_{j} {\pounds_n} K_{ai} 
-W_{ij}^{~\;ab}{\pounds_n} K_{ab}
\bigr)
\Bigr]
=\kappa^2 
S_{ij} -\gamma_{ij}\Lambda
\,,
\label{eq_dynamical}
\end{eqnarray}
respectively, where
\begin{eqnarray}
M_{ijkl}
= &&\!\!\!\!\!
R_{ijkl}-\varepsilon (K_{ik}K_{jl}
-K_{il}K_{jk}),
 \\
M_{ij} 
= &&\!\!\!\!\!
\gamma^{ab}M_{iajb}=R_{ij}
-\varepsilon (KK_{ij}-K_{ia}K^{a}_{~j}), \rule[0mm]{0mm}{5mm}
 \\
M 
= &&\!\!\!\!\!
\gamma^{ab}M_{ab}=R-\varepsilon (K^2-K_{ab}K^{ab}), \rule[0mm]{0mm}{5mm}
  \\
N_{ijk}
= &&\!\!\!\!\!
D_i K_{jk}-D_j K_{ik}, \rule[0mm]{0mm}{5mm}
 \\
N_{i} 
= &&\!\!\!\!\!
\gamma^{ab}N_{aib}=
D_a K_i^{\;a}-D_i K, \rule[0mm]{0mm}{5mm}
\\
H_{ij} 
= &&\!\!\!\!\!
MM_{ij}-2(M_{ia}M^{a}_{~j}
+M^{ab}M_{iajb} )
+M_{iabc}M_{j}^{\;abc} \rule[0mm]{0mm}{5mm}
\nonumber 
\\
&&\!\!\!\!
-2\varepsilon \biggl[-K_{ab}K^{ab}M_{ij}
-{1\over 2} MK_{ia}K^a_{~j} 
+K_{ia}K^a_{~b}M^b_{~j}
+K_{ja}K^a_{~b}M^b_{~i}
+ K^{ac}K_c^{\;b}M_{iajb}
\nonumber 
\\
&& ~~~~~
+N_i N_j-N^a (N_{aij}+N_{aji})
-{1\over 2} N_{abi}N^{ab}_{~~j}
-N_{iab}N^{\;ab}_{j} \biggl]
\nonumber 
\\
&&\!\!\!\!
-{1\over 4}\gamma_{ij}\bigl[
M^2-4M_{ab}M^{ab}+M_{abcd}M^{abcd}\bigr]
\nonumber 
\\
&&\!\!\!\!
-\varepsilon \gamma_{ij}\bigl[K_{ab}K^{ab}M
-2M_{ab}K^{ac}K_{c}^{\;b}
-2N_a N^a 
+N_{abc}N^{abc}\bigr],
  \\
W_{ij}^{\;\;kl}  \rule[0mm]{0mm}{5mm}
= &&\!\!\!\!\!
M\gamma_{ij}\gamma^{kl}-2M_{ij}\gamma^{kl}-2\gamma_{ij}
M^{kl} 
+2M_{iajb}\gamma^{ak}\gamma^{bl}. 
\end{eqnarray}
\end{widetext}
We remark that the terms of ${\pounds_n} K_{ij}$ appear only in the linear form in (\ref{eq_dynamical}). 
This is due to the quasi-linear property of the GB gravity.

\subsection{Cauchy approach}

Using the Bianchi identity, 
Choquet-Bruhat \cite{choquetbruhat1988} showed that the set of equations forms the 
first-class system as the same as in GR, that is 
a space-like hypersurface 
which satisfies the constraints (\ref{eq_Hamiltonian_const}) and
(\ref{eq_momentum_const}) will also satisfies the constraints after the evolution
using (\ref{eq_dynamical}). 

The most standard procedures for following the dynamics of space-time consist from the three steps:  (i) solve the constraint equations (\ref{eq_Hamiltonian_const}) and (\ref{eq_momentum_const}) for 
$(\gamma_{ij}, K_{ij}, \rho_H, J_i)$ on $\Sigma(t=0)$
and prepare them as the initial data, 
(ii) evolve $(\gamma_{ij}, K_{ij}, \rho_H, J_i)$ using (\ref{eq_dynamical}) and the matter equations, and 
(iii) monitor the accuracy of the evolutions 
by checking constraint equations on the evolved $\Sigma(t)$. 

In the case of seeking the ``dynamics" along a space-like direction $\chi$
such as a study of the ``bulk" structure in the brane-world model, 
the above strategy can be switched to the evolution 
in $\chi$-coordinate instead of $t$ (using the set of equations of $\varepsilon=+1$). 
The initial data, in this case, is a time-like hypersurface 
which should satisfy the constraints.  
Such an initial data can be obtained either by solving the dynamics of ``brane" part
or by taking double Wick-rotation after the above step (i), depending on 
the models and motivations. 

In the following sections, we describe a way of solving constraints 
(Sec.~\ref{sec:constraints}), and a way of solving 
evolution equations (Sec.~\ref{sec:evolutions}).

\section{Conformal Approach to solve the Constraints}
\label{sec:constraints}
\subsection{``Conformal Approach"}\label{sec:conformal}

In order to prepare an initial data for dynamical evolution, 
we have to solve two constraints, (\ref{eq_Hamiltonian_const})
and (\ref{eq_momentum_const}).  
The standard approach
is to apply a conformal transformation on the initial hypersurface 
\cite{OMYork74}.  The idea is that introducing a conformal factor 
$\psi$ 
between the initial trial metric 
$\hat{\gamma}_{ij}$ and the solution $\gamma_{ij}$, as 
\begin{equation}
\gamma_{ij}=\psi^{2m}\hat{\gamma}_{ij}, \;\;
\gamma^{ij}=\psi^{-2m}\hat{\gamma}^{ij},
\end{equation}
where $m$ is a constant, 
and solve for $\psi$ so as to the solution satisfies the constraints.

For $N$-dimensional spacetime, Ricci scalar is transformed as
\begin{eqnarray}\label{Ricciscalar_conf}
\!\!\!\!\! R   
\!\!\!\!\! &&= 
 \psi^{-2m} \Bigl\{\hat{R}
-2(N-1)m\psi^{-1}(\hat{D}^a\hat{D}_a \psi) 
\nonumber \\
&& ~~~
+(N-1)\bigl[2-(N-2)m\bigr]m\psi^{-2}(\hat{D} \psi)^2 \Bigr\}, 
\end{eqnarray}
\rule[0mm]{0mm}{0mm}
\begin{eqnarray}\label{Ricci_conf}
\!\!\!\!\! R_{ij} 
\!\!\!\!\! &&= \hat{R}_{ij}
-m\hat{\gamma}_{ij}\psi^{-1}\hat{D}_a\hat{D}^a\psi
\nonumber \\
&& ~~~
-(N-2)m\psi^{-1}\hat{D}_i\hat{D}_j\psi
\nonumber \\
&& ~~~
+(N-2)m(m+1)\psi^{-2}\hat{D}_i\psi \hat{D}_j \psi
\nonumber \\
&& ~~~
-m\bigl[(N-2)m-1\bigr] \psi^{-2}(\hat{D}\psi)^2\hat{\gamma}_{ij}, 
\end{eqnarray}
\begin{eqnarray}\label{Riemann_conf}
\!\!\!\!\! R_{ijkl}   
\!\!\!\!\! &&= 
\psi^{2m}\Bigl\{\hat{R}_{ijkl}
\nonumber \\
&& ~~~
+m\psi^{-1}\hat{\gamma}_{il}\bigl[\hat{D}_j\hat{D}_k\psi-(m+1)\psi^{-1}\hat{D}_j\psi\hat{D}_k\psi\bigr]
\nonumber \\
&& ~~~
-m\psi^{-1}\hat{\gamma}_{ik}\bigl[\hat{D}_j\hat{D}_l\psi-(m+1)\psi^{-1}\hat{D}_j\psi\hat{D}_l\psi\bigr]
\nonumber \\
&& ~~~
+m\psi^{-1}\hat{\gamma}_{jk}\bigl[\hat{D}_i\hat{D}_l\psi-(m+1)\psi^{-1}\hat{D}_i\psi\hat{D}_l\psi\bigr]
\nonumber \\
&& ~~~
-m\psi^{-1}\hat{\gamma}_{jl}\bigl[\hat{D}_i\hat{D}_k\psi-(m+1)\psi^{-1}\hat{D}_i\psi\hat{D}_k\psi\bigr]
\nonumber \\
&& ~~~
+m^2\psi^{-2}(\hat{D}\psi)^2(\hat{\gamma}_{il}\hat{\gamma}_{jk}-\hat{\gamma}_{ik}\hat{\gamma}_{jl})
\Bigr\}.
\end{eqnarray}

Regarding to the extrinsic curvature, we decompose $K_{ij}$ into its
trace part, $K=\gamma^{ij}K_{ij}$, and the traceless part, 
$A_{ij}= K_{ij}-{1 \over N}\gamma_{ij}K$, and assume the conformal 
transformation 
\footnote{In the strict sense this is not the conformal transformation but
just the relation between the values with and without a caret. } as 
\begin{eqnarray}
A_{ij} &=& \psi^{\ell}\hat{A}_{ij}, \;\;
A^{ij} = \psi^{\ell-4m}\hat{A}^{ij}, \label{Aij_conf}\\
K &=& \psi^{\tau}\hat{K},  \label{trK_conf}
\end{eqnarray}
where $\ell$ and $\tau$ are constants. 
For the matter terms, we also assume the relations 
$\rho = \psi^{-p} \hat{\rho}$ and 
$J^i = \psi^{-q} \hat{J}^i$, where $p$ and $q$ are constants, while we 
regard the cosmological constant is common to the both flames, 
$\Lambda = \hat{\Lambda}$.  

Up to here, the powers of conformal transformation, $\ell, m, \tau, p$ and $q$
are not yet specified.  Note that 
in the standard 3-dimensional initial-data construction 
cases, the combination of $m=2$, $\ell=-2$, $\tau=0$, $p=5$ and $q=10$ is preferred since 
this simplifies the equations.  
We also remark that if we chose $\tau=\ell-2m$, then 
the extrinsic curvature can be transformed as $K_{ij}=\psi^{\ell}\hat{K}_{ij}$
and $K^{ij}=\psi^{\ell-4m}\hat{K}^{ij}$.

\subsection{Hamiltonian constraint}
Using these equations, the Hamiltonian constraint equation 
(\ref{eq_Hamiltonian_const}) turns to be
\begin{widetext}
\begin{eqnarray}
&& \!\!\!\!\!\! {2(N-1)m}\hat{D}_a\hat{D}^a\psi
- (N-1)\bigl[2-(N-2)m\bigr]m(\hat{D} \psi)^2 \psi^{-1}
\nonumber \\
&& =
 \hat{R}\psi
-\frac{N-1}{N}\varepsilon\psi^{2m+2\tau+1} \hat{K}^2
+\varepsilon\psi^{-2m+2\ell+1}\hat{A}_{ab}\hat{A}^{ab} 
+2\varepsilon\kappa^2\hat{\rho}\psi^{-p}
-2 \hat{\Lambda} 
+\alpha_{GB} \hat{\Theta}
\psi^{2m+1}.
\label{conf_Hamiltonian1}
\end{eqnarray}
We will show the explicit form of the GB part 
$\hat{\Theta}=M^2-4M_{ab}M^{ab}+M_{abcd}M^{abcd}$ in Appendix ~\ref{app-H}.  At this moment, we observe that (\ref{conf_Hamiltonian1})
can be simplified in the following two ways. 
\begin{itemize} 
\item[(A)] 
If we specify $\tau=\ell-2m$ and $m=2/(N-2)$, 
then (\ref{conf_Hamiltonian1}) becomes \color{black}
\begin{eqnarray}
{4(N-1)\over N-2}\hat{D}_a\hat{D}^a\psi
&=& 
 \hat{R}\psi
 -\varepsilon \psi^{2\ell+1-4/(N-2)}\bigl( \hat{K}^2
 - \hat{K}_{ab}\hat{K}^{ab}\bigr)
+2 \varepsilon \kappa^2\hat{\rho}\psi^{-p}
-2 \hat{\Lambda} 
+\alpha_{GB} \hat{\Theta}
\psi^{1+4/(N-2)}.
\label{conf_Hamiltonian2}
\end{eqnarray}
In the case of the Einstein gravity ($\alpha_{GB}=0$) with $\Lambda=0$, 
the combination $\ell = 2/(N-2)$ and $p=-1$  makes the RHS of
(\ref{conf_Hamiltonian2}) linear.  If we choose $\ell=-2$, which will make the 
momentum constraint simpler as we see later, (\ref{conf_Hamiltonian2}) also remains 
as a simple equation.  
\item[(B)]
If we specify $\tau=0$ and $m=2/(N-2)$, then (\ref{conf_Hamiltonian1}) becomes 
\begin{eqnarray}
{4(N-1)\over N-2}\hat{D}_a\hat{D}^a\psi
&=& 
 \hat{R}\psi
-\varepsilon {N-1 \over N} \psi^{1+4/(N-2)} \hat{K}^2
+\varepsilon 
 \psi^{2\ell+1-4/(N-2)} \hat{A}_{ab}\hat{A}^{ab} 
+2\varepsilon\kappa^2\hat{\rho}\psi^{-p}
-2\hat{\Lambda} 
+\alpha_{GB}  \hat{\Theta}
\psi^{1+4/(N-2)}.
\nonumber \\
\label{conf_Hamiltonian3}
\end{eqnarray}

\end{itemize}

\subsection{Momentum constraint}
In order to express the momentum constraint equation in a tractable form, 
additionally to the variables in Sec.~\ref{sec:conformal}, 
we introduce the transverse traceless part and 
the longitudinal part of $\hat{A}^{ij}$ as
\begin{eqnarray}
&&\hat{D}_j\hat{A}^{ij}_{TT} = 0, \\
&&\hat{A}^{ij}_L=\hat{A}^{ij}-\hat{A}^{ij}_{TT}, 
\end{eqnarray}
respectively. 
Since 
the conformal transformation of the divergence $D_jA_i^{~j}$ becomes
\begin{equation} 
D_jA_i^{~j} =\psi^{\ell-2m} \Bigl\{\hat{D}_j\hat{A}_{i}^{~j}
+\psi^{-1} \bigl[\ell+m(N-2)\bigr] \hat{A}_i^{~j}\hat{D}_j \psi\Bigr\}, 
\label{DA_conformal}
\end{equation}
the momentum constraint, (\ref{eq_momentum_const}), 
can be written as 
\begin{eqnarray}
&&  \hspace{-7mm} 
\psi^{\ell-2m} \hat{D}_a\hat{A}_{i\, L}^{~a}
+\bigl[\ell+(N-2)m\bigr]\psi^{\ell-2m-1}\hat{A}_{i\, L}^{~a}\hat{D}_a\psi
-\frac{N-1}{N}\hat{D}_i (\psi^\tau \hat{K})
+2\alpha_{GB}\hat{\Xi}_i
=\kappa^2 \psi^{2m-q} \hat{J}_i 
\label{conf_momentum}
\end{eqnarray}
\end{widetext}

\color{black}
We will show the explicit form of the GB part 
$\hat{\Xi}_i$ in Appendix ~\ref{app-M},  
meanwhile we proceed the 
standard procedure.  That is, introducing 
 a vector potential $W^i$ for $\hat{A}^{ij}_L$ as
\begin{equation}
\hat{A}^{ij}_L=\hat{D}^iW^j+\hat{D}^jW^i
-{2\over N}\hat{\gamma}^{ij}\hat{D}_kW^k.
\end{equation}
Since the divergence of $\hat{A}^{ij}_L$ becomes
\begin{equation}
\hat{D}_j\hat{A}^{ij}_L
= 
\hat{D}_a\hat{D}^aW^i + {N-2 \over N} \hat{D}^i
\hat{D}_k W^k +\hat{R}^i_{~k} W^k, 
\end{equation}
the momentum constraint, (\ref{conf_momentum}), becomes  
\begin{widetext}
\begin{eqnarray}
&&\!\!\!\!\!\!
\hat{D}_a\hat{D}^aW_i + {N-2 \over N} \hat{D}_i
\hat{D}_k W^k +\hat{R}_{ik} W^k
+\psi^{-1}\bigl[\ell+(N-2)m\bigr]
\Bigl( \hat{D}^aW^b+\hat{D}^bW^a
-{2\over N}\hat{\gamma}^{ab}\hat{D}_kW^k \Bigr) \hat{\gamma}_{bi}
\hat{D}_a\psi
\nonumber \\
&&
-\psi^{2m-\ell} \frac{N-1}{N}\hat{D}_i(\psi^\tau \hat{K})
+\psi^{2m-\ell} 2\alpha_{GB}\hat{\Xi}_i
=\kappa^2  \psi^{4m-\ell-q} \hat{J}_i 
\label{conf_momentum1}
\end{eqnarray}
Similarly to the case of the Hamiltonian constraint equation, we consider the two cases. 
\begin{itemize}
\item[(A)]
If we 
specify $\tau=\ell-2m$ and $m=2/(N-2)$, then (\ref{conf_momentum1}) becomes 
\begin{eqnarray}
&&\!\!\!\!\!\!
\hat{D}_a\hat{D}^aW_i + {N-2 \over N} \hat{D}_i
\hat{D}_k W^k +\hat{R}_{ik} W^k
+\psi^{-1}(\ell+2)
\Bigl( \hat{D}^aW^b+\hat{D}^bW^a
-{2\over N}\hat{\gamma}^{ab}\hat{D}_kW^k \Bigr) \hat{\gamma}_{bi}
\hat{D}_a\psi
\nonumber \\
&&
-\frac{N-1}{N}\left[ \Bigl(\ell - {4\over N-2}\Bigr)(\hat{D}_i\psi)\hat{K} + \hat{D}_i\hat{K} \right]
+\psi^{-\ell+4/(N-2)} 2\alpha_{GB}\hat{\Xi}_i
=\kappa^2  \psi^{8/(N-2)-\ell-q} \hat{J}_i 
\label{conf_momentum2}
\end{eqnarray}
In the case of the Einstein gravity ($\alpha_{GB}=0$), 
the choice of $\ell=-2$ cancels the mixing term between $\psi$ and $W^i$.
Further, 
when $\alpha_{GB}=0$, we have a chance to make two constraint equations, 
(\ref{conf_Hamiltonian2}) and (\ref{conf_momentum2}), decouple by assuming 
$\hat{K}=0$ and $q=8/(N-2)+2$.  However, when $\alpha_{GB}\neq 0$, this 
decoupling feature is no longer available, since the term $\hat{\Xi}_i$ includes 
$\psi$-related terms as we see in eq. (\ref{appM_caseA_xi}) in Appendix~\ref{app-M}. 
\item[(B)]
If we specify $\tau=0$ and $m=2/(N-2)$, then (\ref{conf_momentum1}) becomes 
\begin{eqnarray}
&&\!\!\!\!\!\!
\hat{D}_a\hat{D}^aW_i + {N-2 \over N} \hat{D}_i
\hat{D}_k W^k +\hat{R}_{ik} W^k
+\psi^{-1}(\ell+2)
\bigl[ \hat{D}^aW^b+\hat{D}^bW^a
-{2\over N}\hat{\gamma}^{ab}\hat{D}_kW^k \bigr] \hat{\gamma}_{bi}
\hat{D}_a\psi
\nonumber \\
&&
-\psi^{4/(N-2)-\ell} \frac{N-1}{N}\hat{D}_i\hat{K}
+\psi^{4/(N-2)-\ell} 2\alpha_{GB}\hat{\Xi}_i
=\kappa^2  \psi^{8/(N-2)-\ell-q} \hat{J}_i 
\label{conf_momentum3}
\end{eqnarray}
For the Einstein gravity ($\alpha_{GB}=0$),  
the choice of $\ell=-2$ again cancels the mixing term between $\psi$ and $W^i$. 
The decoupling feature between (\ref{conf_Hamiltonian3}) and (\ref{conf_momentum3})
is available when $\alpha_{GB}=0$, $\hat{K}=$const. and $q=8/(N-2)+2$.
However, when $\alpha_{GB}\neq 0$, this 
decoupling feature is no longer available, since the term $\hat{\Xi}_i$ includes 
$\psi$-related terms as we see in eq. (\ref{appM_caseB_xi}) in Appendix~\ref{app-M}. 
\end{itemize}
\end{widetext}

\subsection{Procedures}
For the readers' convenience, we summarize the above procedure briefly. 
The initial data, $(\gamma_{ij}, K_{ij}, \rho, J^i)$, 
can be constructed by solving the Hamiltonian constraint, (\ref{eq_Hamiltonian_const}), 
and the momentum constraint equations, (\ref{eq_momentum_const}). 
This can be done by the following steps. 
\begin{enumerate}
\item Give the initial assumption (trial values) for 
$\hat{\gamma}_{ij}, K,$ $\hat{A}_{ij}^{TT}$ and 
$\hat{\rho},~\hat{J}$. 
\item Solve (\ref{conf_Hamiltonian1}) and (\ref{conf_momentum1}) for 
$\psi$ and $W^i$ by fixing the exponent $\ell, m, \tau, p$ and $q$. 
\item By the following inverse conformal transformations, 
\begin{eqnarray}
\gamma_{ij} &=& \psi^{2m} \hat{\gamma}_{ij},  \\
K_{ij} &=& \psi^{\ell}[\hat{A}_{ij}^{TT}+
\hat{D}_iW_j+\hat{D}_jW_i
-{2\over N}\hat{\gamma}_{ij}\hat{D}_kW^k] \nonumber \\
&&+{1 \over N}\psi^{2m+\tau} \hat{\gamma}_{ij} \hat{K},  \\
\rho &=& \psi^{-p} \hat{\rho},  \\
J^i &=& \psi^{-q} \hat{J}^i,
\end{eqnarray}
we obtain the solution $\gamma_{ij},~K_{ij},~\rho,~J^i$, which satisfy the constraints. 
\end{enumerate}

\subsection{Momentarily static case}
The most easiest construction of an initial data may be under the assumption
of the momentarily static (or time-symmetric) situation, 
$K_{ij}=J_i=0$.  
In such a case, the momentum constraint
becomes trivial, and the 
Hamiltonian constraint (\ref{conf_Hamiltonian1}) can be reduced as 
\begin{widetext}
\begin{eqnarray}
\!\!\!\!\!\!
&&{2(N-1)m}\hat{D}_a\hat{D}^a\psi
- (N-1)\bigl[2-(N-2)m\bigr]m(\hat{D} \psi)^2 \psi^{-1}
= \hat{R}\psi
+2\varepsilon\kappa^2\hat{\rho}\psi^{-p}
-2\hat{\Lambda}
+\alpha_{GB} \hat{\Theta}
\psi^{2m+1}, \color{black}
\label{conf_Hamiltonian1_timesym}
\end{eqnarray}
where
\begin{eqnarray}
&&\!\!\!\!\!\!
\hat{\Theta}=
(N-3)m \psi^{-4m}
\biggl\{
4(N-2) m\psi^{-2}\Bigl[(\hat{D}_a\hat{D}^a\psi)^2
-(\hat{D}_{a}\hat{D}_{b}\psi)(\hat{D}^{a}\hat{D}^{b}\psi)\Bigr]
\nonumber \\
&& ~~~~~~
-4\psi^{-1}\Bigl[\hat{R}
-(N-2) \bigl[(N-3)m-2\bigr]m\psi^{-2}
(\hat{D} \psi)^2 \Bigr]\hat{D}_a\hat{D}^a\psi 
\nonumber \\
&& ~~~~~~
+8\psi^{-1}\Bigl[\hat{R}^{ab}
+(N-2)m(m+1)\psi^{-2}\hat{D}^{a}\psi\hat{D}^{b}\psi
\Bigr]\hat{D}_a\hat{D}_b\psi
\nonumber \\
&& ~~~~~~
+(N-1)_2m^2\bigl[(N-4)m-4\bigr] \psi^{-4}(\hat{D} \psi)^4
-2\psi^{-2}\bigl[(N-4)m-2\bigr] 
(\hat{D} \psi)^2\hat{R}
\nonumber \\
&& ~~~~~~
-8(m+1)\psi^{-2}\hat{R}^{ab} \hat{D}_a\psi\hat{D}_b\psi
\biggr\}
+\psi^{-4m}\hat{R}_{GB}, 
\end{eqnarray}
\end{widetext}
where 
\begin{equation}
(D-n)_m={(D-n)! \over (D-m-1)!}=(D-n)(D-n-1)\cdots(D-m),
\end{equation}
and  $\hat{R}_{GB}=\hat{R}^2-4\hat{R}_{ab}\hat{R}^{ab}+\hat{R}_{abcd}\hat{R}^{abcd}$. 

When $N=3$, $\hat{\Theta}$ simply becomes $\hat{\Theta}=\psi^{-4m}\hat{R}_{GB}$ so that 
(\ref{conf_Hamiltonian1_timesym}) will be reduced as
\begin{equation}
8\hat{D}_a\hat{D}^a\psi
= \hat{R}\psi
+2\varepsilon\kappa^2\hat{\rho}\psi^{-p}
-2\hat{\Lambda}
+\alpha_{GB} \hat{R}_{GB}
\psi^{-3},
\label{conf_Hamiltonian1_timesym2}
\end{equation}
for the choice of $m=2$. 

However, in general $N$, it is hard to find out an appropriate 
$m$ which simplifies the equation, even if  
 $\hat{\gamma}_{ij}$ is taken to be the flat space-time that reduces $\hat{\Theta}$ as 
\begin{eqnarray}
\label{hamiltonian-time-sym}
&&\!\!\!\!\!\!
\hat{\Theta}=
(N-3)m^2 \psi^{-4m-2}
\nonumber \\
&& ~
\times \biggl\{
4(N-2) \Bigl[(\hat{D}_a\hat{D}^a\psi)^2
-(\hat{D}_{a}\hat{D}_{b}\psi)(\hat{D}^{a}\hat{D}^{b}\psi)\Bigr]
\nonumber \\
&& ~~~~~~
+4
(N-2) \bigl[(N-3)m-2\bigr]\psi^{-1}(\hat{D}\psi)^2
\hat{D}_a\hat{D}^a\psi
\nonumber \\
&& ~~~~~~
+8
(N-2)(m+1)\psi^{-1}(\hat{D}^{a}\psi)(\hat{D}^{b}\psi)
\hat{D}_a\hat{D}_b\psi
\nonumber \\
&& ~~~~~~
+(N-1)_2m\bigl[(N-4)m-4\bigr] \psi^{-2}(\hat{D}\psi)^4
\biggr\}. 
\end{eqnarray}

Roughly speaking, Eq.~(\ref{hamiltonian-time-sym})  is a quadratic equation with respect to the second order derivative of $\psi$, which means that there are two roots in general when a set of trial values $\hat{\gamma}_{ij}$ and $\hat{\rho}$ on the hypersurface are given. The meaning of the existence of two solutions is more clearly understood by assuming $p=0$ and $q=0$ in the conformal transformation of the matter field. In this case, two different $\gamma_{ij}$ are obtained through the different conformal transformations even for the same matter field distributions.

The non-uniqueness of the solution is due to  the higher curvature combination of the GB terms. The existence of two solutions can be also seen for the black hole solutions in the Einstein-GB theory\cite{Boulware,Torii_2}. In the black hole case, the gravitational equations
reduce to a quadratic equation under the suitable choice of a metric function. The two roots of the equation correspond to the different black hole solutions.
In the $\alpha_{GB} \to 0$ limit, one of the black hole solutions approaches to the solutions in GR while the metric component of the other solution diverges and there is no counterpart in GR. Hence the former is classified into the GR branch and the latter into the GB branch. In some mass parameter regions of the black hole, the metric components of these branches coincide at the certain radius. It is a multiple root of the field equation.
Solving the field equations beyond this radius, we find that the metric becomes imaginary and this region is physically irrelevant. The spacetime becomes singular at this radius and it is called branch singularity, which is a new ingredients of GB gravity.

When we solve the Hamiltonian constraint equation (\ref{hamiltonian-time-sym}), the similar situation may occur. The conformal factor $\psi$ becomes imaginary and the solution $\gamma_{ij}$ is also imaginary in some regions on the hypersurface.  The boundary of this region corresponds to the singularity. The situation is same also in the non-time symmetric case.

\section{Dynamical equations}
\label{sec:evolutions}
\subsection{Dynamical equations in General Relativity}
Equations in GR are obtained by putting $\alpha_{GB}=0$. 
The evolution equations of ($N$+1)-dimensional ADM formulation
are presented in \cite{ndimCP}, but here we generalize them to the
set of equations both for spacelike and timelike foliations. 

Firstly we rewrite the dynamical equations (\ref{eq_dynamical}) by introducing the metric components as 
\begin{eqnarray}
ds^2 && \!\!\!\!\!
= \varepsilon\alpha^2 {(dx^0)}^2 +\gamma_{ij}(dx^i+\beta^i dx^0) (dx^j+\beta^j dx^0)
\nonumber \\ 
&&  \!\!\!\!\!
=(\varepsilon\alpha^2+\beta_{a}\beta^{a}){(dx^0)}^2
+2\beta_{i} dx^i dx^0
+\gamma_{ij}dx^i dx^j ,
\nonumber \\ 
\label{metric}
\end{eqnarray}
where $\alpha$ and $\beta^i$
are the lapse and shift functions, respectively.
The components of the normal vector, then,  are
\begin{equation}
n_{\mu} =(\varepsilon \alpha, \;0, \;\cdots, \; 0), \;\;\;\;
n^{\mu} =\frac{1}{\alpha}(1, \; -\beta^{i}).
\end{equation}
In the matrix from, the $(N+1)$- and $N$- metrics 
are expressed as
\begin{equation}
g_{\mu\nu} =\left(
\begin{array}{cc}
\varepsilon\alpha^2+\beta_{a}\beta^{a} & \;\; \beta_{j} \\
\beta_{i} & \;\; \gamma_{ij}
\end{array}
\right),
\end{equation}
\begin{equation}
g^{\mu\nu} =\frac{\varepsilon}{\alpha^2}\left(
\begin{array}{cc}
1 & \;\; -\beta^{i} \\
-\beta^{j} & \;\; \varepsilon\alpha^2\gamma^{ij}+\beta^{i}\beta^{j}
\end{array}
\right),
\end{equation}
and 
\begin{equation}
\gamma_{\mu\nu} =\left(
\begin{array}{cc}
\beta_{a}\beta^{a} & \;\; \beta_{j} \\
\beta_{i} & \;\; \gamma_{ij}
\end{array}
\right),
\;\;\;
\gamma^{\mu\nu} =\left(
\begin{array}{cc}
\:0 &  0 \\
\:0 &  \;\gamma^{ij}
\end{array}
\right). 
\end{equation}

Trace of Eq. (\ref{eq_dynamical}) minus half of Eq.
(\ref{eq_Hamiltonian_const}) gives
\begin{equation}
\gamma^{ab}\pounds_{n} K_{ab}=-\frac{\varepsilon}{2}M-K_{ab}K^{ab}
-\frac{\varepsilon\kappa^{2}}{N-1}{\cal T}
+\varepsilon \Lambda.
\label{traceLie}
\end{equation}  
This is equivalent to the trace of Einstein equation (\ref{Einstein}) with the projection (\ref{Riemann_nn}). Substituting Eq.~(\ref{traceLie}) into Eq.~(\ref{eq_dynamical}), we find
\begin{eqnarray}
&&{\pounds_n} K_{ij}=-\varepsilon M_{ij}-K_{ia}K^a_{~j}
\nonumber
\\
&& ~~~~
+\varepsilon\kappa^{2}\biggl(S_{ij}
-\frac{1}{N-1}{\cal T}
\gamma_{ij}\biggr)
+{2\varepsilon\over N-1}\gamma_{ij}\Lambda.
\label{dynamical-gr}
\end{eqnarray}

The extrinsic curvature and its Lie derivative are expressed as
\begin{eqnarray}
K_{ij} ={ 1\over 2 \alpha}
\left( -  \partial_{0} \gamma_{ij} + D_j \beta_ {i} + D_i \beta_{j}
\right),
\label{ex-curv-2}
\end{eqnarray}  
\begin{eqnarray}
&& \!\!\!\!\!
\pounds_{n}K_{ij}=
\frac{1}{\alpha}\bigl(\partial_{0} K_{ij} + D_iD_j \alpha
-\beta^{a}D_{a}K_{ij}
\nonumber \\
&& ~~~~~~~~~~~~~~
-K_{aj}D_{i}\beta^{a}-K_{ai}D_{j}\beta^{a}\bigr)
\label{Lie-K}
\end{eqnarray}
respectively.

With the metric components (\ref{metric}), the equations (\ref{ex-curv-2}) and (\ref{dynamical-gr})  are
\begin{eqnarray}
&&\partial_{0} \gamma_{ij}=-2 \alpha K_{ij} + D_j \beta_{i} + D_i \beta_{j},
\\
&& \!\!\!
\partial_{0} K_{ij}  = 
-\alpha\varepsilon M_{ij}-\alpha K_{ia}K^a_{~j} - D_iD_j \alpha
+\beta^a  (D_a K_{ij})
\nonumber  \\
&& ~~~~~~~~~~
+(D_j\beta^a)K_{ia}
+(D_i\beta^a)K_{aj}
\rule[0mm]{0mm}{5mm}.
\nonumber  \\
&&~~~~~~~~~~
- \alpha\varepsilon\kappa^2 \biggl( S_{ij} 
-  {1\over  N-1}{\cal T}
\gamma_{ij}  \biggr)
\rule[0mm]{0mm}{6mm}
+{2\alpha\varepsilon  \over N-1} \gamma_{ij}\Lambda.
\nonumber  \\
\label{KijevoNdim}
\end{eqnarray}

If we have matter, we need to evolve them together with metric.
The dynamical equations for matter terms can be derived from the
conservation
equation,
$\nabla^\mu {\cal T}_{\mu\nu}=0$.

\subsection{Dynamical equations in Gauss-Bonnet gravity}
With the Gauss-Bonnet terms, the evolution equation, 
(\ref{eq_dynamical}), cannot be expressed explicitly for each ${\pounds_n} K_{ij}$. 
That is, (\ref{eq_dynamical}) is rewritten as
\begin{widetext}
\begin{eqnarray}
&&
(1+2\alpha_{GB}M){\pounds_n} K_{ij}
-(\gamma_{ij}\gamma^{ab}+2\alpha_{GB}W_{ij}^{~\;ab}){\pounds_n} K_{ab}
-8\alpha_{GB}M^{\;\;a}_{(i} {\pounds_n} K_{|a| j)}
\rule[0mm]{0mm}{4mm}
\nonumber \\
&&
\;\;\;\;\;\;\;
=-\varepsilon\biggl(M_{ij}-{1\over2}M\gamma_{ij}\biggr)
-K_{ia}K^a_{~j}+\gamma_{ij}K_{ab}K^{ab}
+\varepsilon\kappa^2 S_{ij}
-\varepsilon\gamma_{ij}\Lambda
-2\varepsilon\alpha_{GB}H_{ij}\,, \label{dyn_GB}
\rule[0mm]{0mm}{5mm}
\end{eqnarray}
\end{widetext}
and the second and third terms in r.h.s include 
the  linearly-coupled terms between ${\pounds_n} K_{ij}$. 
Therefore, in an actual simulation, we have to extract each 
evolution equation of $K_{ij}$ using a matrix form of 
Eq.~(\ref{dyn_GB}) like
\begin{equation}
\label{matrixform}
{\mathbf k}=A{\mathbf k}+{\mathbf b}
\end{equation}
where ${\mathbf k}=({\pounds_n} K_{11}, {\pounds_n} K_{12}, \cdots, {\pounds_n} K_{NN})^{T}$ and $A,\: {\mathbf b}$ are appropriate matrix and vector derived from Eq.~(\ref{dyn_GB}). 

The procedure of the inverting the matrix $(1-A)$ is technically available,
but the invertibility of the matrix is not generally guaranteed at this moment.
In the case of the standard ADM foliation in 4-dimensional Einstein equations,
the continuity of the time evolutions depends on the models and the choice of
gauge conditions for the lapse function and shift vectors.  If the combination
is not appropriate, then the foliation hits the singularity which stops
the evolution.  The similar obstacle may exist also for the Gauss-Bonnet
gravity.  We expect that in the most cases Eq.~(\ref{matrixform}) is invertible for
$K_{ij}$ but we cannot deny the pathological cases which depend on the models
and gauge conditions.  Such a study must be done together with actual numerical
integrations in the future.

We obtain, however, the similar form of equation by introducing the $(N+1)$-dimensional Weyl curvature\cite{Shiromizu,Maeda}. It is useful for the discussion of the brane-world cosmology although it is not written by only the $N$-dimensional quantities, and we can not adopt it as the evolution equation directly.
Trace of Eq.~(\ref{eq_dynamical}) minus half of Eq.~(\ref{eq_Hamiltonian_const}) gives
\begin{eqnarray}
&& \!
(N-1)\biggl[\frac{\varepsilon}2 M +K_{ab}K^{ab}+h^{ab}\pounds_{n} K_{ab}\biggr]
\nonumber \\
&& \;\;\;\;
+2(n-3)\alpha_{GB}\biggl[\frac{\varepsilon}{4}\bigl(M^{2}-4M_{ab}M^{ab}+M_{abcd}M^{abcd}\bigr)
\nonumber \\
&& \;\;\;\;
+MK_{ab}K^{ab}-2K^{i}_{\;j}K^{j}_{\;k}M^{k}_{\;\;i}
-2N_{a}N^{a}
\nonumber \\
&& \;\;\;\;
+N_{abc}N^{abc}+Mh^{ab}\pounds_{n} K_{ab}
-2M^{ab}\pounds_{n} K_{ab}\biggr]
\nonumber \\
&& \;\;\;\;
=-\varepsilon\kappa^{2}{\cal T}+\varepsilon \Lambda
\label{traceLieGB}
\end{eqnarray}  
By the last term in the l.h.s of this equation, the term of the Lie derivative cannot
be expressed by other term explicitly. 

Let us rewrite the dynamical equation (\ref{eq_dynamical}) in a different form.
From Eqs. (\ref{Riemann_nn}), (\ref{5DRicci}) and (\ref{5Dscalar}) with
the decomposition of the Riemann tensor as
\begin{eqnarray}
&&
{\cal R}_{\mu\nu\rho\sigma}
=\frac{2}{N-1}\left(g_{\mu [\rho}{\cal R}_{\sigma]\nu}
-g_{\nu [ \rho}{\cal R}_{\sigma] \mu}\right)
\nonumber \\
&& ~~~~~~~~~~~~
-\frac{2}{N(N-1)} g_{\mu [\rho} g_{\sigma ]\nu}{\cal R}
+{\cal C}_{\mu\nu\rho\sigma}\, ,
\label{Riemann}
\end{eqnarray}
where ${\cal C}_{\mu\nu\rho\sigma}$ is the (N+1)-dimensional Weyl curvature,
we find
\begin{eqnarray}
&& \!\!\!
{\pounds_n} K_{ij}=\varepsilon \frac{N-1}{N-2} E_{ij}
+ \frac{\varepsilon}{N-2}\Biggl( M_{ij}-\frac{1}{N}M\gamma_{ij}\biggr)
\nonumber \\
&& ~~~~~~~~
-K_{ia}K^{a}_{~j}
+\frac{1}{N} \gamma_{ij}K_{ab}K^{ab}
+\frac{1}{N} \gamma_{ij} \gamma^{ab}{\pounds_n}K_{ab}\,,
\nonumber \\
\label{LKMN1}
\end{eqnarray}
where
\begin{equation}
E_{ij} := {\cal C}_{\mu\nu\rho\sigma}~n^{\mu} n^{\rho}
\gamma_{~i}^{\nu} \gamma_{~j}^{\sigma} \,.
\label{Edef}
\end{equation}
However, because  Eq. (\ref{LKMN1}) is a trace free equation,
${\pounds_n} K_{\mu\nu}$ cannot be fixed  by Eq. (\ref{LKMN1}).
Inserting 
Eq. (\ref{LKMN1}) into Eq. (\ref{traceLieGB}),
we find
\begin{eqnarray}
h^{ab}{\pounds_n}K_{ab}
=&& \!\!\!\!\!\! -\frac{\varepsilon}{2}M
-K_{ab}K^{ab}
\nonumber \\
&&\!\!\!\!\!\! 
-\frac{\varepsilon}{U}(\kappa^2{\cal T}-\Lambda)
+\frac{2\varepsilon(N-3)\alpha_{GB}}{U} I,
\label{traceLKMN}
\end{eqnarray}
where
\begin{eqnarray}
&& 
U=N-1+\frac{2(N-2)(N-3)}{N}\alpha_{GB} M,
\\
&& 
I=
\frac{N-6}{4(N-2)}M^2
+\frac{N}{N-2}M_{ab}M^{ab}
+\frac14 M_{abcd}M^{abcd}
\rule[0mm]{0mm}{8mm}
\nonumber \\
&& 
~~~~~
-\varepsilon \biggl[N_a N^a +4 N_{abc}N^{abc}
-\frac{2(N-1)}{N-2}M_{ab}E^{ab}\biggr].
\rule[0mm]{0mm}{8mm}
\nonumber \\
\end{eqnarray}
From Eq.  (\ref{LKMN1})  with Eq. (\ref{traceLKMN}), we then find
\begin{eqnarray}
&& \!\!\!\!\!\!\!\!\!\!\!\!\!
{\pounds_n}K_{ij}
=\frac{N-1}{N-2} E_{ij}
+\frac{\varepsilon}{N-2}\left(M_{ij}-\frac12 M\gamma_{ij}\right)
\nonumber \\
&&~~~~
-K_{ia}K^{a}_{\;j}
-\frac{\varepsilon}{NU}(\kappa^2{\cal T} - \Lambda) \gamma_{ij}
\nonumber \\
&&~~~~
-\frac{2\varepsilon (N-3)\alpha_{GB}}{NU}I \gamma_{ij}.
\label{LKMN2}
\end{eqnarray}
Since $E_{ij}$ is $(N+1)$-dimensional quantity, ${\pounds_n}K_{ij}$ can not be evaluated by the valuables on $N$-dimensional hypersurface with this equation.
This means that the Eq.~(\ref{LKMN2}) can not be used for the full dynamics as we mentioned.
For example, however, for the Friedmann brane-world model, where the constant time slice of the timelike hypersurface is homogeneous and isotropic, $E_{ij}$ can be written using quantities on the hypersurface. For such limited cases where the term $E_{ij}$ can be evaluated on the hypersurface, the Eq.~(\ref{LKMN2}) is useful for simplifyng the situation.

\section{Discussion}
\label{sec:discussion}

With the aim of numerical investigations of spacetime 
dynamics in  higher-dimensional and/or higher-curvature gravity
models, we presented the basic equations of the 
Einstein-Gauss-Bonnet gravity theory.

We show the $(N+1)$-dimensional decomposition of the basic equations, 
in order to treat the space-time as a Cauchy problem.  
With the aim of investigations of bulk spacetime in recent brane-world models, 
we also prepared the equations both for timelike and spacelike foliations. 
The equations can be separated to the constraints (the Hamiltonian constraint and the momentum constraint) and the evolution equations.  

Two constraints should be solved for constructing an initial data. 
By showing the conformally-transformed constraint equations, 
we discussed how the constraints can be 
simplified by tuning the powers of conformal factors. 
If we have Gauss-Bonnet terms, however, the equations still remain in 
a complicated style.  

For the evolution equations, we find that ${\pounds_n} K_{ij}$ components
are coupled. However, this mixture is only up to the linear order due to  
the quasi-linear property of the Gauss-Bonnet terms, so that the equations 
can be in a treatable form in numerics. 

We are now developing our numerical code and hope to present some 
results elsewhere near future. 

\appendix 
\section{GB part of Hamiltonian constraint equation}
\label{app-H}
In Eq.~(\ref{conf_Hamiltonian1}), the GB part of the Hamiltonian constraint equation $\hat{\Theta}=M^2-4M_{ab}M^{ab}+M_{abcd}M^{abcd}$ is not written explicitly. It becomes as
\begin{widetext}
\begin{eqnarray}
&&\!\!\!\!\!\!
\hat{\Theta}=
(N-3)m\psi^{-4m}\biggl\{
4(N-2) m\psi^{-2}\Bigl[(\hat{D}_a\hat{D}^a\psi)^2
-\hat{D}_{a}\hat{D}_{b}\psi\hat{D}^{a}\hat{D}^{b}\psi\Bigr]
\nonumber \\
&& ~~~~~
-4\psi^{-1}\Bigl[\hat{M}
-(N-2) \bigl[(N-3)m-2\bigr]m\psi^{-2}\hat{D}_{a}\psi\hat{D}^{a}\psi
\Bigr]\hat{D}_a\hat{D}^a\psi
\nonumber \\
&& ~~~~~
+8\psi^{-1}\Bigl[\hat{M}^{ab}
+(N-2)m(m+1)\psi^{-2}\hat{D}^{a}\psi\hat{D}^{b}\psi
\Bigr]\hat{D}_a\hat{D}_b\psi
\nonumber \\
&& ~~~~~
+(N-1)_2 m^2\bigl[(N-4)m-4\bigr] \psi^{-4}(\hat{D}_a \psi \hat{D}^a \psi)^2
-2\psi^{-2}\bigl[(N-4)m-2\bigr]\hat{M}
\hat{D}_c\psi\hat{D}^c\psi
\nonumber \\
&& ~~~~~
-8(m+1)\psi^{-2}
\hat{M}^{ab}
 \hat{D}_a\psi\hat{D}_b\psi
\biggr\}
+\psi^{-4m}(\hat{\Upsilon}^2-4\hat{\Upsilon}_{ab}\hat{\Upsilon}^{ab}+\hat{\Upsilon}_{abcd}\hat{\Upsilon}^{abcd}),
\end{eqnarray}
where
\begin{eqnarray}
&&
\hat{\Upsilon}=\hat{R}
-\varepsilon \biggl[\frac{N-1}{N}\psi^{2m+2\tau} \hat{K}^2
-\psi^{2\ell-2m}\hat{A}_{ab} \hat{A}^{ab}\biggr],
\\
&&
\hat{\Upsilon}_{ij}=\hat{R}_{ij}
-\varepsilon \biggl[
\frac{N-1}{N^2}\psi^{2m+2\tau}\hat{\gamma}_{ij}\hat{K}^2
+\frac{N-2}{N}\psi^{\ell+\tau}\hat{K}\hat{A}_{ij}
-\psi^{2\ell-2m}\hat{A}_{ia}\hat{A}^{a}_{\:\:j}\biggr],
\\
&&
\hat{\Upsilon}_{ijkl}=\hat{R}_{ijkl}
-\varepsilon\biggl[
\frac{1}{N^2}\psi^{2m+2\tau}(\hat{\gamma}_{ik}\hat{\gamma}_{jl}-\hat{\gamma}_{il}\hat{\gamma}_{jk})\hat{K}^2
+\frac{1}{N}\psi^{\ell+\tau}(\hat{A}_{ik}\hat{\gamma}_{jl}-\hat{A}_{il}\hat{\gamma}_{jk}
+\hat{A}_{jl}\hat{\gamma}_{ik}-\hat{A}_{jk}\hat{\gamma}_{il})
\nonumber \\
&& \hspace{30mm}
+\psi^{2\ell-2m}(\hat{A}_{ik}\hat{A}_{jl}-\hat{A}_{il}\hat{A}_{jk})\biggr].
\end{eqnarray}

When $\tau=\ell-2m$ and $m=2/(N-2)$, which corresponds to the case (A) [Eq.~(\ref{conf_Hamiltonian2})] 
\begin{eqnarray}
\label{Theta_2}
&&\!\!\!\!\!\!
\hat{\Theta}=
\frac{2(N-3)}{N-2}\psi^{-8/(N-2)}\biggl\{
8\psi^{-2}\Bigl[(\hat{D}_a\hat{D}^a\psi)^2
-\hat{D}_{a}\hat{D}_{b}\psi\hat{D}^{a}\hat{D}^{b}\psi\Bigr]
\nonumber \\
&& ~~~~~
-4\psi^{-1}\bigl(\hat{M}+2\psi^{-2}\hat{D}_{a}\psi\hat{D}^{a}\psi
\bigr)\hat{D}_a\hat{D}^a\psi
+8\psi^{-1}\Bigl(\hat{M}^{ab}
+\frac{2N}{N-2}\psi^{-2}\hat{D}^{a}\psi\hat{D}^{b}\psi
\Bigr)\hat{D}_a\hat{D}_b\psi
\nonumber \\
&& ~~~~~
-\frac{8N(N-1)}{(N-2)^2}\psi^{-4}(\hat{D}_a \psi \hat{D}^a \psi)^2
+\frac{8}{N-2}\psi^{-2}\hat{M}\hat{D}_c\psi\hat{D}^c\psi
-\frac{8N}{N-2}\psi^{-2}\hat{M}^{ab}\hat{D}_a\psi\hat{D}_b\psi
\biggr\}
\nonumber \\
&& ~~~~~
+\psi^{-8/(N-2)}(\hat{\Upsilon}^2-4\hat{\Upsilon}_{ab}\hat{\Upsilon}^{ab}+\hat{\Upsilon}_{abcd}\hat{\Upsilon}^{abcd}),
\end{eqnarray}
where
\begin{eqnarray}
&&
\hat{\Upsilon}=\hat{R}-\varepsilon \psi^{2\ell-4/(N-2)} (\hat{K}^2-\hat{K}_{ab} \hat{K}^{ab}),
\\
&&
\hat{\Upsilon}_{ij}=\hat{R}_{ij}
-\varepsilon \psi^{2\ell-4/(N-2)}(\hat{K}\hat{K}_{ij}-\hat{K}_{ia}\hat{K}^{a}_{\:\:j}),
\\
&&
\hat{\Upsilon}_{ijkl}=\hat{R}_{ijkl}
-\varepsilon \psi^{2\ell-4/(N-2)}(\hat{K}_{ik}\hat{K}_{jl}-\hat{K}_{il}\hat{K}_{jk}).
\end{eqnarray}

When $\tau=0$, $m=2/(N-2)$, which corresponds to the case (B) [Eq.~(\ref{conf_Hamiltonian3})], $\Theta$ is expressed as Eq.~(\ref{Theta_2}) and
\begin{eqnarray}
&&
\hat{\Upsilon}=\hat{R}
-\varepsilon \biggl[\frac{N-1}{N}\psi^{4/(N-2)} \hat{K}^2
-\psi^{2\ell-4/(N-2)}\hat{A}_{ab} \hat{A}^{ab}\biggr],
\\
&&
\hat{\Upsilon}_{ij}=\hat{R}_{ij}
-\varepsilon \biggl[
\frac{N-1}{N^2}\psi^{4/(N-2)}\hat{\gamma}_{ij}\hat{K}^2
+\frac{N-2}{N}\psi^{\ell}\hat{K}\hat{A}_{ij}
-\psi^{2\ell-4/(N-2)}\hat{A}_{ia}\hat{A}^{a}_{\:\:j}\biggr],
\\
&&
\hat{\Upsilon}_{ijkl}=\hat{R}_{ijkl}
-\varepsilon\biggl[
\frac{1}{N^2}\psi^{4/(N-2)}(\hat{\gamma}_{ik}\hat{\gamma}_{jl}-\hat{\gamma}_{il}\hat{\gamma}_{jk})\hat{K}^2
+\frac{1}{N}\psi^{\ell}(\hat{A}_{ik}\hat{\gamma}_{jl}-\hat{A}_{il}\hat{\gamma}_{jk}
+\hat{A}_{jl}\hat{\gamma}_{ik}-\hat{A}_{jk}\hat{\gamma}_{il})
\nonumber \\
&& \hspace{30mm}
+\psi^{2\ell-4/(N-2)}(\hat{A}_{ik}\hat{A}_{jl}-\hat{A}_{il}\hat{A}_{jk})\biggr],
\end{eqnarray}

\section{GB part of Momentum constraint equation}
\label{app-M}
In Eq.~(\ref{conf_momentum1}), the GB part of the Hamiltonian constraint equation $\hat{\Xi}_i$ is not written explicitly. It becomes as
\begin{eqnarray}
&&  \hspace{-12mm} 
\Xi_i=
\psi^{\ell-4m}\biggl\{
\hat{R}
-2(N-3)m\psi^{-1}\hat{D}_b\hat{D}^b\psi
-(N-3)m\bigr[(N-4)m+2\bigr]\psi^{-2}\hat{D}_b\psi\hat{D}^b\psi
\nonumber \\ 
&& 
-\frac{N^2-3N+4}{N^2}\varepsilon \psi^{+2m+2\tau}\hat{K}^2
-\varepsilon \psi^{2\ell-2m}\hat{A}_{bc}\hat{A}^{bc}
\biggr\}
\hat{D}_a \hat{A}^a_{~i}
\nonumber \\ 
&& \hspace{-5mm} 
+\psi^{\ell-4m}\biggl\{
-2\hat{R}^b_{~i}
+2(N-3)m\psi^{-1}\hat{D}^b\hat{D}_i\psi
-2(N-3)m(m+1)\psi^{-2}\hat{D}_i\psi\hat{D}^b\psi
\nonumber \\ 
&& 
+\frac{2(N-3)}{N}\varepsilon \psi^{\ell+\tau}\hat{K}\hat{A}_{i}^{~b}
-2\varepsilon \psi^{2\ell-2m}\hat{A}_i^{~c}\hat{A}_c^{~b}
\biggr\}
\hat{D}_a\hat{A}^a_{~b}
\nonumber \\ 
&& \hspace{-5mm} 
+\psi^{\ell-4m}\biggl\{
2\hat{R}^{ab}
-2(N-3)m\psi^{-1}\hat{D}^b\hat{D}^a \psi
-2(N-1)m(m+1)\psi^{-2}\hat{D}^a\psi\hat{D}^b \psi
\nonumber \\ 
&& 
-\frac{2(N-3)}{N}\varepsilon \psi^{\ell+\tau}\hat{K}\hat{A}^{ab}
+2\varepsilon\psi^{2\ell-2m}\hat{A}^a_{~c}\hat{A}^{cb}
\biggl\}
\bigl(\hat{D}_i\hat{A}_{ab}-\hat{D}_a\hat{A}_{ib} \bigr)
\nonumber \\ 
&& \hspace{-5mm} 
+2\varepsilon \psi^{3\ell-6m}\hat{A}_i^{~a}\hat{A}^{bc} 
\bigl(\hat{D}_a\hat{A}_{bc}-\hat{D}_b\hat{A}_{ac} \bigr)
+{\cal R}_i+{\cal D}_i
+{\cal A}^{(1)}\hat{D}_i\psi
+{\cal A}^{(2)}\hat{D}_i\hat{K}
+{\cal A}^{(3)}\hat{D}_a\psi\hat{A}^a_{~i}
\nonumber \\ 
&& \hspace{-5mm} 
-\frac{2(N-2)_3}{N^2}\varepsilon
\psi^{\ell-2m+2\tau}\hat{K}(\hat{D}_a\hat{K}+\tau\psi^{-1}\hat{K}\hat{D}_a\psi)\hat{A}^a_{~i}
\nonumber \\ 
&& \hspace{-5mm} 
+\frac{2(N-3)}{N}\bigl[(N-4)m+2\ell+\tau\bigr]\varepsilon
\psi^{2\ell-4m+\tau-1}\hat{K}\hat{D}_b\psi\hat{A}^b_{~a}\hat{A}^a_{~i}
\nonumber \\ 
&& \hspace{-5mm} 
+\frac{2(N-3)}{N}\varepsilon
\psi^{2\ell-4m+\tau}\hat{D}_b\hat{K}\hat{A}^b_{~a}\hat{A}^a_{~i}
-2\bigl[(N-6)m+3\ell\bigr]\varepsilon
\psi^{3\ell-6m-1}\hat{D}_c\psi\hat{A}^c_{~b}\hat{A}^b_{~a}\hat{A}^a_{~i},
\end{eqnarray}
where
\begin{eqnarray}
&&  \hspace{-10mm} 
{\cal R}_i=
\biggl\{\bigr[(N-3)m+\ell\bigr]\psi^{\ell-4m-1}\hat{A}_i^{~a}\hat{D}_a\psi
-\frac{N-3}{N}\psi^{-2m+\tau}(\hat{D}_i\hat{K}+\tau\psi^{-1}\hat{K}\hat{D}_i\psi)
\biggr\}\hat{R}
\nonumber \\ 
&& 
+\biggl\{\frac{2(N-3)}{N}\tau\psi^{-2m+\tau-1}\hat{K}\hat{D}_a\psi
+\frac{2(N-3)}{N}\psi^{-2m+\tau}\hat{D}_a\hat{K}
-2\bigl[(N-3)m+\ell\bigr]\psi^{\ell-4m-1}\hat{A}_a^{~b}\hat{D}_b\psi
\biggr\}\hat{R}_i^{~a}
\nonumber \\ 
&& 
-2(m-\ell)\psi^{\ell-4m-1}(\hat{A}_{ab}\hat{D}_i\psi-\hat{A}_{ib}\hat{D}_a\psi)
\hat{R}^{ab}
+2(m-\ell)\psi^{\ell-4m-1}\hat{D}_a\psi\hat{A}_{bc}\hat{R}_i^{~cab},
\end{eqnarray}
\begin{eqnarray}
&&  \hspace{-10mm} 
{\cal D}_i=
\biggl\{\frac{N^2-8N+11}{N}m\psi^{-2m+\tau-1}
(\hat{D}_i\hat{K}+\tau\psi^{-1}\hat{K}\hat{D}_i\psi)
\rule[0mm]{75mm}{0mm}
\nonumber \\ 
&& ~~~~~ 
-2m\bigl[(N^2-6N+7)m+(N-3)\ell\bigr]
\psi^{\ell-4m-2}\hat{D}_b\psi\hat{A}^b_{~i}
\biggr\}\hat{D}_a\hat{D}^a\psi
\nonumber \\ 
&& 
-\biggl\{\frac{2(N-2)_3}{N}m\psi^{-2m+\tau-1}
(\hat{D}_a\hat{K}+\tau\psi^{-1}\hat{K}\hat{D}_a\psi)
\nonumber \\ 
&& ~~~~~ 
-2m\bigl[(N^2-4N+5)m+(N-2)(\ell-2)\bigr]
\psi^{\ell-4m-2}\hat{D}_b\psi\hat{A}^b_{~a}
\biggr\}\hat{D}^a\hat{D}_i\psi
\nonumber \\ 
&& 
+2(N-3)m(m-\ell)\psi^{\ell-4m-2}(\hat{A}_{ab}\hat{D}_i\psi-\hat{A}_{ia}\hat{D}_b\psi)
\hat{D}^b\hat{D}^a\psi,
\end{eqnarray}
\begin{eqnarray}
&&  \hspace{-1mm}  
{\cal A}^{(1)}
=2\biggl\{
-\frac{N-2}{N}m(m+1)\psi^{-2m+2\tau-2}
(\hat{D}_a\hat{K}+\tau\psi^{-1}\hat{K}\hat{D}_a\psi)
(\hat{D}^a\hat{K}+\tau\psi^{-1}\hat{K}\hat{D}^a\psi)
\nonumber \\ 
&& ~~~~~ ~~~~~~~
+\frac{(N-2)^2}{N}m(m+1)\psi^{-2m+\tau-2}\hat{D}_a\hat{K}\hat{D}^a\psi
+\frac{N-2}{2N}m\bigl[(N^2-4N+5)m+2\bigr]\tau\psi^{-2m+\tau-3}\hat{K}\hat{D}_a\psi\hat{D}^a\psi
\nonumber \\ 
&& ~~~~~ ~~~~~~~
-(N-2)_3m^2(m+1)\psi^{\ell-4m-3}\hat{D}^a\psi\hat{D}^b\psi\hat{A}_{ab}
+\frac{N-3}{N}(m-\ell-\tau)\varepsilon\psi^{2\ell-4m+\tau-1}\hat{K}\hat{A}_{ab}\hat{A}^{ab}
\nonumber \\ 
&& ~~~~~ ~~~~~~~
-(m-\ell)\varepsilon\psi^{3\ell-6m-1}\hat{A}_a^{~b}\hat{A}_b^{~c}\hat{A}_c^{~a}
\biggr\},
\end{eqnarray}
\begin{eqnarray}
&&  \hspace{-10mm}  
{\cal A}^{(2)}
=\frac{1}{N}\biggl\{
(N-2)_3m\bigl[(N-3)m-2\bigr]\psi^{-2m+\tau-2}\hat{D}_a\psi\hat{D}^a\psi
-\frac{(N-1)_2(N+1)}{N^2}\varepsilon\psi^{3\tau}\hat{K}^2
\rule[0mm]{25mm}{0mm}
\nonumber \\ 
&& ~~~~~
-(N-3)\varepsilon\psi^{2\ell-4m+\tau}\hat{A}_{ab}\hat{A}^{ab}
\biggr\},
\end{eqnarray}
\begin{eqnarray}
&&  \hspace{-10mm}  
{\cal A}^{(3)}
=-
m\bigl[(N-2)^2(N-5)m^2+(N-2)_3(\ell-2)m
+(N-1)(3\ell-2)\bigr]\psi^{\ell-4m-3}\hat{D}_a\psi\hat{D}^a\psi
\rule[0mm]{17mm}{0mm}
\nonumber \\ 
&& ~
-\frac{1}{N^2}\bigl[(N-1)(N^2-8)m
+(N^2-N+2)\ell\bigr]\varepsilon\psi^{\ell-2m+2\tau-1}\hat{K}^2
\nonumber \\ 
&&  ~
+\bigl[(N-6)m+3\ell\bigr]\varepsilon\psi^{3\ell-6m-1}\hat{A}_{ab}\hat{A}^{ab}.
\end{eqnarray}

When $\tau=\ell-2m$ and $m=2/(N-2)$, which corresponds to the case (A) [Eq.~(\ref{conf_momentum2})], 
\begin{eqnarray}
&&  \hspace{-12mm} 
\Xi_i=
\psi^{\ell-8/(N-2)}\biggl\{
\hat{R}
-\frac{4(N-3)}{N-2}\psi^{-1}\hat{D}_b\hat{D}^b\psi
-\frac{4(N-3)^2}{N-2}\psi^{-2}\hat{D}_b\psi\hat{D}^b\psi
\nonumber \\ 
&& ~~~~~ ~~~~~~~
-\frac{N^2-3N+4}{N^2}\varepsilon \psi^{2\ell-4/(N-2)}\hat{K}^2
-\varepsilon \psi^{2\ell-4/(N-2)}\hat{A}_{bc}\hat{A}^{bc}
\biggr\}
\hat{D}_a \hat{A}^a_{~i}
\nonumber \\ 
&& \hspace{-5mm} 
+\psi^{\ell-8/(N-2)}\biggl\{
-2\hat{R}^b_{~i}
+\frac{4(N-3)}{N-2}\psi^{-1}\hat{D}^b\hat{D}_i\psi
-\frac{4N(N-3)}{(N-2)^2}\psi^{-2}\hat{D}_i\psi\hat{D}^b\psi
\nonumber \\ 
&&  ~~~~~ ~~~~~~~
+\frac{2(N-3)}{N}\varepsilon \psi^{2\ell-4/(N-2)}\hat{K}\hat{A}_{i}^{~b}
-2\varepsilon \psi^{2\ell-4/(N-2)}\hat{A}_i^{~c}\hat{A}_c^{~b}
\biggr\}
\hat{D}_a\hat{A}^a_{~b}
\nonumber \\ 
&& \hspace{-5mm} 
+\psi^{\ell-8/(N-2)}\biggl\{
2\hat{R}^{ab}
-\frac{4(N-3)}{N-2}\psi^{-1}\hat{D}^b\hat{D}^a \psi
-\frac{4N(N-1)}{(N-2)^2}\psi^{-2}\hat{D}^a\psi\hat{D}^b \psi
\nonumber \\ 
&&  ~~~~~ ~~~~~~~
-\frac{2(N-3)}{N}\varepsilon \psi^{2\ell-4/(N-2)}\hat{K}\hat{A}^{ab}
+2\varepsilon\psi^{2\ell-4/(N-2)}\hat{A}^a_{~c}\hat{A}^{cb}
\biggl\}
\bigl(\hat{D}_i\hat{A}_{ab}-\hat{D}_a\hat{A}_{ib} \bigr)
\nonumber \\ 
&& \hspace{-5mm} 
+2\varepsilon \psi^{3\ell-12/(N-2)}\hat{A}_i^{~a}\hat{A}^{bc} 
\bigl(\hat{D}_a\hat{A}_{bc}-\hat{D}_b\hat{A}_{ac} \bigr)
+{\cal R}_i+{\cal D}_i
+{\cal A}^{(1)}\hat{D}_i\psi
+{\cal A}^{(2)}\hat{D}_i\hat{K}
+{\cal A}^{(3)}\hat{D}_a\psi\hat{A}^a_{~i}
\nonumber \\ 
&& \hspace{-5mm} 
-\frac{2(N-2)_3}{N^2}\varepsilon
\psi^{3\ell-12/(N-2)}\hat{K}\biggl[\hat{D}_a\hat{K}+\Bigl(\ell-\frac{4}{N-2}\Bigr)\psi^{-1}\hat{K}\hat{D}_a\psi\biggr]\hat{A}^a_{~i}
\nonumber \\ 
&& \hspace{-5mm} 
+\frac{2(N-3)}{N}\biggl[3\ell+\frac{2(N-6)}{N-2}\biggr]\varepsilon
\psi^{3\ell-12/(N-2)-1}\hat{K}\hat{D}_b\psi\hat{A}^b_{~a}\hat{A}^a_{~i}
\nonumber \\ 
&& \hspace{-5mm} 
+\frac{2(N-3)}{N}\varepsilon
\psi^{3\ell-12/(N-2)}\hat{D}_b\hat{K}\hat{A}^b_{~a}\hat{A}^a_{~i}
-2\biggl[3\ell+\frac{2(N-6)}{N-2}\biggr]\varepsilon
\psi^{3\ell-12/(N-2)-1}\hat{D}_c\psi\hat{A}^c_{~b}\hat{A}^b_{~a}\hat{A}^a_{~i},
\label{appM_caseA_xi}
\end{eqnarray}
where
\begin{eqnarray}
&&  \hspace{-10mm} 
{\cal R}_i=
\psi^{\ell-8/(N-2)}\biggl\{\Bigr[\ell+\frac{2(N-3)}{N-2}\Bigr]\psi^{-1}\hat{A}_i^{~a}\hat{D}_a\psi
-\frac{N-3}{N}\biggl[\hat{D}_i\hat{K}+\Bigl(\ell-\frac{4}{N-2}\Bigr)\psi^{-1}\hat{K}\hat{D}_i\psi\biggr]
\biggr\}\hat{R}
\nonumber \\ 
&& 
+\psi^{\ell-8/(N-2)}\biggl\{\frac{2(N-3)}{N}\Bigl(\ell-\frac{4}{N-2}\Bigr)\psi^{-1}\hat{K}\hat{D}_a\psi
+\frac{2(N-3)}{N}\hat{D}_a\hat{K}
-2\biggl[\ell+\frac{2(N-3)}{N-2}\biggr]\psi^{-1}\hat{A}_a^{~b}\hat{D}_b\psi
\biggr\}\hat{R}_i^{~a}
\nonumber \\ 
&&  
+2\Bigl(\ell-\frac{2}{N-2}\Bigr)\psi^{\ell-8/(N-2)-1}(\hat{A}_{ab}\hat{D}_i\psi-\hat{A}_{ib}\hat{D}_a\psi)
\hat{R}^{ab}
-2\Bigl(\ell-\frac{2}{N-2}\Bigr)\psi^{\ell-8/(N-2)-1}\hat{D}_a\psi\hat{A}_{bc}\hat{R}_i^{~cab},
\end{eqnarray}
\begin{eqnarray}
&&  \hspace{-10mm} 
{\cal D}_i=
\psi^{\ell-8/(N-2)}\biggl\{\frac{2(N^2-8N+11)}{N(N-2)}\psi^{-1}
\biggl[\hat{D}_i\hat{K}+\Bigl(\ell-\frac{4}{N-2}\Bigr)\psi^{-1}\hat{K}\hat{D}_i\psi\biggr]
\rule[0mm]{53mm}{0mm}
\nonumber \\ 
&& ~~~~~ ~~~~~~~
-\frac{4}{N-2}\biggl[(N-3)\ell+\frac{2(N^2-6N+7)}{N-2}\biggr]
\psi^{-2}\hat{D}_b\psi\hat{A}^b_{~i}
\biggr\}\hat{D}_a\hat{D}^a\psi
\nonumber \\ 
&& 
-\psi^{\ell-8/(N-2)}\biggl\{\frac{4(N-3)}{N}\psi^{-1}
\biggl[\hat{D}_a\hat{K}+\Bigl(\ell-\frac{4}{N-2}\Bigr)\psi^{-1}\hat{K}\hat{D}_a\psi\biggr]
\nonumber \\ 
&& ~~~~~ ~~~~~~~
-4\biggl[\ell+\frac{6}{(N-2)^2}\biggr]
\psi^{-2}\hat{D}_b\psi\hat{A}^b_{~a}
\biggr\}\hat{D}^a\hat{D}_i\psi
\nonumber \\ 
&& 
-\frac{4(N-3)}{N-2}\Bigl(\ell-\frac{2}{N-2}\Bigr)\psi^{\ell-8/(N-2)-2}(\hat{A}_{ab}\hat{D}_i\psi-\hat{A}_{ia}\hat{D}_b\psi)
\hat{D}^b\hat{D}^a\psi,
\end{eqnarray}
\begin{eqnarray}
&&  \hspace{-10mm}  
{\cal A}^{(1)}
=2\psi^{\ell-8}\biggl\{
-\frac{2}{N-2}\psi^{\ell-4/(N-2)-2}
\biggl[\hat{D}_a\hat{K}+\Bigl(\ell-\frac{4}{N-2}\Bigr)\psi^{-1}\hat{K}\hat{D}_a\psi\biggr]
\biggl[\hat{D}^a\hat{K}+\Bigl(\ell-\frac{4}{N-2}\Bigr)\psi^{-1}\hat{K}\hat{D}^a\psi\biggr]
\rule[0mm]{3mm}{0mm}
\nonumber \\ 
&& ~~~~~ ~~~~~~~
+2\psi^{-2}\hat{D}_a\hat{K}\hat{D}^a\psi
+\frac{2(N^2-3n+3)}{N(N-2)}\Bigl(\ell-\frac{4}{N-2}\Bigr)\psi^{-3}\hat{K}\hat{D}_a\psi\hat{D}^a\psi
\nonumber \\ 
&& ~~~~~ ~~~~~~~
-\frac{4N(N-3)}{(N-2)^2}\psi^{-3}\hat{D}^a\psi\hat{D}^b\psi\hat{A}_{ab}
+\frac{2(N-3)}{N}\Bigl(\ell-\frac{1}{N-2}\Bigr)\varepsilon\psi^{\ell-4/(N-2)-1}\hat{K}\hat{A}_{ab}\hat{A}^{ab}
\nonumber \\ 
&& ~~~~~ ~~~~~~~
+\Bigl(\ell-\frac{2}{N-2}\Bigr)\varepsilon\psi^{\ell-4/(N-2)-1}\hat{A}_a^{~b}\hat{A}_b^{~c}\hat{A}_c^{~a}
\biggr\},
\end{eqnarray}
\begin{eqnarray}
&&  \hspace{-10mm}  
{\cal A}^{(2)}
=\frac{1}{N}\psi^{\ell-8/(N-2)}\biggl\{
-\frac{4(N-3)}{N-2}\psi^{-2}\hat{D}_a\psi\hat{D}^a\psi
\rule[0mm]{89mm}{0mm}
\nonumber \\ 
&& ~~~~~ ~~~~~~~
-\frac{(N-1)_2(N+1)}{N^2}\varepsilon\psi^{2\ell-4/(N-2)}\hat{K}^2
-(N-3)\varepsilon\psi^{2\ell-4/(N-2)}\hat{A}_{ab}\hat{A}^{ab}
\biggr\},
\end{eqnarray}
\begin{eqnarray}
&&  \hspace{-10mm}  
{\cal A}^{(3)}
=-\psi^{\ell-8/(N-2)}\biggl\{
\frac{4}{N-2}\bigl[(N-3)\ell-6\bigr]\psi^{-3}\hat{D}_a\psi\hat{D}^a\psi
\rule[0mm]{75mm}{0mm}
\nonumber \\ 
&& ~~~~~ ~~~~~~~
+\frac{1}{N^2}\biggl[(N^2-N+2)\ell+\frac{2(N-1)}{(N-2)(N^2-8)}\biggr]\varepsilon\psi^{2\ell-4-1}\hat{K}^2
\nonumber \\ 
&& ~~~~~ ~~~~~~~
-\biggl[3\ell+\frac{2(N-6)}{N-2}\biggr]\varepsilon\psi^{2\ell-4/(N-2)-1}\hat{A}_{ab}\hat{A}^{ab}\biggr\}.
\end{eqnarray}

When $\tau=0$ and $m=2/(N-2)$, which corresponds to the case (B) [Eq.~(\ref{conf_momentum3})], 
\begin{eqnarray}
&&  \hspace{-12mm} 
\hat{\Xi}_i=
\psi^{\ell-8/(N-2)}\biggl\{
\hat{R}
-\frac{4(N-3)}{N-2}\psi^{-1}\hat{D}_b\hat{D}^b\psi
-\frac{8(N-3)^2}{(N-2)^2}\psi^{-2}\hat{D}_b\psi\hat{D}^b\psi
\nonumber \\ 
&& ~~~~~
-\frac{N^2-3N+4}{N^2}\varepsilon \psi^{4/(N-2)}\hat{K}^2
-\varepsilon \psi^{2\ell-4/(N-2)}\hat{A}_{bc}\hat{A}^{bc}
\biggr\}
\hat{D}_a \hat{A}^a_{~i}
\nonumber \\ 
&& \hspace{-5mm} 
+\psi^{\ell-8/(N-2)}\biggl\{
-2\hat{R}^b_{~i}
+\frac{4(N-3)}{N-2}\psi^{-1}\hat{D}^b\hat{D}_i\psi
-\frac{4N(N-3)}{(N-2)^2}\psi^{-2}\hat{D}_i\psi\hat{D}^b\psi
\nonumber \\ 
&& ~~~~~
+\frac{2(N-3)}{N}\varepsilon \psi^{\ell}\hat{K}\hat{A}_{i}^{~b}
-2\varepsilon \psi^{2\ell-4/(N-2)}\hat{A}_i^{~c}\hat{A}_c^{~b}
\biggr\}
\hat{D}_a\hat{A}^a_{~b}
\nonumber \\ 
&& \hspace{-5mm} 
+\psi^{\ell-8/(N-2)}\biggl\{
2\hat{R}^{ab}
-\frac{4(N-3)}{N-2}\psi^{-1}\hat{D}^b\hat{D}^a \psi
-\frac{4N(N-1)}{(N-2)^2}\psi^{-2}\hat{D}^a\psi\hat{D}^b \psi
\nonumber \\ 
&& ~~~~~
-\frac{2(N-3)}{N}\varepsilon \psi^{\ell}\hat{K}\hat{A}^{ab}
+2\varepsilon\psi^{2\ell-4/(N-2)}\hat{A}^a_{~c}\hat{A}^{cb}
\biggl\}
\bigl(\hat{D}_i\hat{A}_{ab}-\hat{D}_a\hat{A}_{ib} \bigr)
\nonumber \\ 
&& \hspace{-5mm} 
+2\varepsilon \psi^{3\ell-12/(N-2)}\hat{A}_i^{~a}\hat{A}^{bc} 
\bigl(\hat{D}_a\hat{A}_{bc}-\hat{D}_b\hat{A}_{ac} \bigr)
+{\cal R}_i+{\cal D}_i
+{\cal A}^{(1)}\hat{D}_i\psi
+{\cal A}^{(2)}\hat{D}_i\hat{K}
+{\cal A}^{(3)}\hat{D}_a\psi\hat{A}^a_{~i}
\nonumber \\ 
&& \hspace{-5mm} 
-\frac{2(N-2)_3}{N^2}\varepsilon
\psi^{\ell-4/(N-2)}\hat{K}\hat{D}_a\hat{K}\hat{A}^a_{~i}
+\frac{4(N-3)}{N}\Bigl(\ell+\frac{N-4}{N-2}\Bigr)\varepsilon
\psi^{2\ell-8/(N-2)-1}\hat{K}\hat{D}_b\psi\hat{A}^b_{~a}\hat{A}^a_{~i}
\nonumber \\ 
&& \hspace{-5mm} 
+\frac{2(N-3)}{N}\varepsilon
\psi^{2\ell-8/(N-2)}\hat{D}_b\hat{K}\hat{A}^b_{~a}\hat{A}^a_{~i}
-2\biggl[3\ell+\frac{2(N-6)}{N-2}\biggr]\varepsilon
\psi^{3\ell-12/(N-2)-1}\hat{D}_c\psi\hat{A}^c_{~b}\hat{A}^b_{~a}\hat{A}^a_{~i},
\label{appM_caseB_xi}
\end{eqnarray}
where
\begin{eqnarray}
&&  \hspace{-10mm} 
{\cal R}_i=
\biggl\{\Bigr[\ell+\frac{2(N-3)}{N-2}\Bigr]\psi^{\ell-8/(N-2)-1}\hat{A}_i^{~a}\hat{D}_a\psi
-\frac{N-3}{N}\psi^{-4/(N-2)}\hat{D}_i\hat{K}
\biggr\}\hat{R}
\rule[0mm]{43mm}{0mm}
\nonumber \\ 
&& 
+2\biggl\{\frac{(N-3)}{N}\psi^{-4/(N-2)}\hat{D}_a\hat{K}
-\Bigr[\ell+\frac{2(N-3)}{N-2}\Bigr]\psi^{\ell-8/(N-2)-1}\hat{A}_a^{~b}\hat{D}_b\psi
\biggr\}\hat{R}_i^{~a}
\nonumber \\ 
&& 
+2\Bigl(\ell-\frac{2}{N-2}\Bigr)\psi^{\ell-8/(N-2)-1}
\Bigl\{(\hat{A}_{ab}\hat{D}_i\psi-\hat{A}_{ib}\hat{D}_a\psi)\hat{R}^{ab}
-\hat{D}_a\psi\hat{A}_{bc}\hat{R}_i^{~cab}\Bigr\},
\end{eqnarray}
\begin{eqnarray}
&&  \hspace{-10mm} 
{\cal D}_i=
\frac{2}{N-2}\biggl\{\frac{N^2-8N+11}{N}\psi^{-4/(N-2)-1}\hat{D}_i\hat{K}
\rule[0mm]{83mm}{0mm}
\nonumber \\ 
&& ~~~~~~ ~~~~~~~
-2\biggl[(N-3)\ell+\frac{2(N^2-6N+7)}{N-2}\biggr]
\psi^{\ell-8/(N-2)-2}\hat{D}_b\psi\hat{A}^b_{~i}
\biggr\}\hat{D}_a\hat{D}^a\psi
\nonumber \\ 
&& 
-4\biggl\{\frac{N-3}{N}\psi^{-4/(N-2)-1}\hat{D}_a\hat{K}
-\biggl[\ell+\frac{6}{(N-2)^2}\biggr]
\psi^{\ell-8/(N-2)-2}\hat{D}_b\psi\hat{A}^b_{~a}
\biggr\}\hat{D}^a\hat{D}_i\psi
\nonumber \\ 
&& 
-\frac{4(N-3)}{N-2}\Bigl(\ell-\frac{2}{N-2}\Bigr)\psi^{\ell-8/(N-2)-2}(\hat{A}_{ab}\hat{D}_i\psi-\hat{A}_{ia}\hat{D}_b\psi)
\hat{D}^b\hat{D}^a\psi,
\end{eqnarray}
\begin{eqnarray}
&&  \hspace{-10mm} 
{\cal A}^{(1)}
=
-\frac{4}{N-2}\psi^{-4/(N-2)-2}\hat{D}_a\hat{K}\hat{D}^a\hat{K}
+4\psi^{-4/(N-2)-2}\hat{D}_a\hat{K}\hat{D}^a\psi
\rule[0mm]{56mm}{0mm}
\nonumber \\ 
&& 
-\frac{8N(N-3)}{(N-2)^2}\psi^{\ell-8/(N-2)-3}\hat{D}^a\psi\hat{D}^b\psi\hat{A}_{ab}
-\frac{2(N-3)}{N}\Bigl(\ell-\frac{2}{N-2}\Bigr)\varepsilon\psi^{2\ell-8/(N-2)-1}\hat{K}\hat{A}_{ab}\hat{A}^{ab}
\nonumber \\ 
&& 
+2\Bigl(\ell-\frac{2}{N-2}\Bigr)\varepsilon\psi^{3\ell-12/(N-2)-1}\hat{A}_a^{~b}\hat{A}_b^{~c}\hat{A}_c^{~a},
\end{eqnarray}
\begin{eqnarray}
&&  \hspace{-10mm} 
{\cal A}^{(2)}
=-\frac{1}{N}\biggl\{
2(N-3)\psi^{-4/(N-2)-2}\hat{D}_a\psi\hat{D}^a\psi
+\frac{(N-1)_2(N+1)}{N^2}\varepsilon\hat{K}^2
+(N-3)\varepsilon\psi^{2\ell-8/(N-2)}\hat{A}_{ab}\hat{A}^{ab}
\biggr\},
\end{eqnarray}
\begin{eqnarray}
&&  \hspace{-10mm}  
{\cal A}^{(3)}
=-\frac{4}{N-2}\bigl[(N-3)\ell-6\bigr]\psi^{\ell-8/(N-2)-3}\hat{D}_a\psi\hat{D}^a\psi
\rule[0mm]{75mm}{0mm}
\nonumber \\ 
&& 
-\frac{1}{N^2}\biggl[(N^2-N+2)\ell+\frac{2(N-1)(N^2-8)}{N-2}\biggr]\varepsilon\psi^{\ell-4/(N-2)-1}\hat{K}^2
\nonumber \\ 
&& 
+\biggl[3\ell+\frac{2(N-6)}{N-2}\biggr]\varepsilon\psi^{3\ell-12/(N-2)-1}\hat{A}_{ab}\hat{A}^{ab}.%
\end{eqnarray}
\end{widetext}

\section*{Acknowledgments}

H.S. was partially supported by the Special Research Fund
(Project No. 4244) of the Osaka Institute of Technology, Faculty of Information Science
and Technology. 

\vspace{10mm}

\end{document}